\documentclass[12pt]{article}
\usepackage{epsfig,epsf}

\begin{document}


\begin{center}
{\Large\bf Rigidity of Scattering Lengths and Travelling Times for Disjoint Unions of 
Convex Bodies}
\end{center}

\begin{center}
{\sc Lyle Noakes and Luchezar Stoyanov}
\end{center}

\newcommand{\R}{{\sf I\hspace{-.15em}R}}
\def\C{{\bf C}}
\def\e{\emptyset}
\def\ds{\partial S}
\def\cq{\overline{Q}}
\def\sn{{\bf S}^{n-1}}
\def\sd{{\bf S}^{d-1}}
\def\hg{\Gamma}
\def\ssd{{\sd}\times {\sd}}
\def\do{\partial \Omega}
\def\dk{\partial K}
\def\dl{\partial L}
\def\tts{\tilde{S}}
\def\tsigma{\tilde{\sigma}}
\def\ts{\tilde{s}}
\def\ttp{\tilde{p}}
\def\tu{\tilde{u}}
\def\tU{\tilde{U}}
\def\hS{\hat{S}}
\def\hs{\hat{s}}
\def\hp{\hat{p}}
\def\dr{\frac{\partial r}{\partial x_1}}
\def\ll{{\cal L}}
\def\hm{\hat{M}}
\def\pp{{\cal P}}
\def\T{{\cal T}}
\def\tt{\tilde{t}}
\def\ds{\partial S}
\def\ss{{\cal S}}
\def\vv{{\cal V}}
\def\aa{{\cal A}}
\def\bb{{\cal B}}
\def\nn{{\cal N}}
\def\dd{{\cal D}}
\def\ee{{\cal E}}
\def\uu{{\cal U}}
\def\rr{{\cal R}}
\def\kk{{\cal K}}
\def\cc{C_0^{\infty}}
\def\ot{(\omega,\theta)}
\def\oto{(\omega_0,\theta_0)}
\def\toto{(\tilde{\omega}_0, \tilde{\theta}_0)}
\def\tx{\tilde{x}}
\def\txi{\tilde{\xi}}
\def\got{\gamma(\omega,\theta)}
\def\ggot{\gamma'(\omega,\theta)}
\def\omo{(\omega,\omega)}
\def\du{\partial u_i}
\def\dfu{\frac{\partial f}{\partial u_i}}
\def\dou{\frac{\partial \omega}{\partial u_i}}
\def\dttu{\frac{\partial \theta}{\partial u_i}}
\def\ooo{(\omega_0,-\omega_0)}
\def\otp{(\omega',\theta')}
\def\oo{{\cal O}}
\def\pr{{\rm pr}}
\def\cK{\hat{K}}
\def\cL{\hat{L}}
\def\ff{{\cal F}}
\def\fk{{\cal F}^{(K)}}
\def\fl{{\cal F}^{(L)}}
\def\kkr{{\cal K}^{\mbox{reg}}}
\def\kkro{{\cal K}_0^{\mbox{reg}}}
\def\trapk{\mbox{\rm Trap}(\Omega_K)}
\def\trapl{\mbox{\rm Trap}(\Omega_L)}
\def\oock{\stackrel{\circ}{\Omega}_{\cK}}
\def\oocl{\stackrel{\circ}{\Omega}_{\cL}}
\def\sok{S_b^*(\Omega_K)\setminus \trapk}
\def\sol{S_b^*(\Omega_L)\setminus \trapl}
\def\sbok{S_b^*(\Omega_K)}
\def\sbol{S_b^*(\Omega_L)}
\def\hr{\hat{\rho}}
\def\G{{\cal G}}
\def\tg{\tilde{\gamma}}
\def\Vo{\mbox{Vol}}
\def\dkso{\partial K^{(\infty)}}
\def\dlso{\partial L^{(\infty)}}
\def\gg{{\cal g}}
\def\sl{{\cal SL}}
\def\tkn{\mbox{\rm Trap}^{(n)}(\dk)}
\def\tln{\mbox{\rm Trap}^{(n)}(\dl)}
\def\wn{{\cal WN}}
\def\i{{\bf i}}
\def\te{{\cal T}^{(ext)}}
\def\ndk{{\cal N}_b^*(\dk)}
\def\ndl{{\cal N}_b^*(\dl)}
\def\dr{\frac{\partial r}{\partial z_1}}
\def\endofproof{{\rule{6pt}{6pt}}}
\def\Box{\endofproof}
\def\su{S^*(\R^n\setminus U)}
\def\gk{\gamma_K}
\def\gl{\gamma_L}
\def\la{\left\langle}
\def\ra{\right\rangle}
\def\kfin{\kk_0^{({\mbox{\footnotesize\rm fin}})}} 
\def\dt{\dot{T}}
\def\ep{\epsilon}
\def\kfi{\kk^{({\mbox{\footnotesize\rm fin}})}} 
\def\stk{\Sigma_3^{(K)}}
\def\stl{\Sigma_3^{(L)}}
\def\SU{S^*(\R^n\setminus U)}
\def\tkm{\tt_k^{(m)}}
\def\ukm{U_k^{(m)}}
\def\Pkm{\Psi_k^{(m)}}
\def\nkm{N_k^{(m)}}
\def\di{\displaystyle}
\def\nk{N^{(K)}}
\def\nl{N^{(L)}}
\def\mk{M^{(K)}}
\def\ml{M^{(L)}}
\def\ep{\epsilon}
\def\Gk{G^k}
\def\uk{U^{(K)}}
\def\dist{\mbox{\rm dist}}
\def\diam{\mbox{\rm diam}}
\def\con{\mbox{\rm const}}
\def\bs{\bigskip}
\def\ms{\medskip}
\def\SK{S^*_{K}(S_0)}
\def\SL{S^*_{L}(S_0)}

\def\te{\tilde{e}}
\def\tx{\tilde{x}}
\def\ty{\tilde{y}}
\def\tv{\tilde{v}}
\def\grad{\; \mbox{\rm grad} \,}
\def\sock{S^*_+(S_0)\setminus \trapk}
\def\socl{S^*_+(S_0)\setminus \trapl}

\def\trapkk{\mbox{\rm Trap}^+(K)}
\def\tX{\widetilde{X}}
\def\tF{\widetilde{F}}
\def\tY{\widetilde{Y}}
\def\tZ{\widetilde{Z}}
\def\hY{\widehat{Y}}

\footnotesize

\noindent
{\sc Abstract.} Obstacles $K$ and $L$ in $\R^d$ ($d\geq 2$) are considered  that are finite disjoint 
unions of strictly convex domains with $C^3$ boundaries. We show that if $K$ and $L$ have (almost) 
the same scattering length spectrum, or (almost) the same travelling times, then $K = L$. 

\normalsize



\section{Introduction}
\renewcommand{\theequation}{\arabic{section}.\arabic{equation}}

Let $K$ be a compact subset of ${\R}^d$ ($d \geq 2$) with $C^{3}$ boundary $\partial K$ such that 
$\Omega_K = \overline{{\R}^d\setminus K}$ is connected. A {\it scattering ray} $\gamma$ in 
$\Omega_K$ is an unbounded in both directions generalized geodesic (in the sense of Melrose and 
Sj\"ostrand \cite{kn:MS1}, \cite{kn:MS2}). If $K$ is a finite disjoint union of convex domains,
then the scattering rays in $\Omega_K$ are simply billiard trajectories with finitely many common
points with $\dk$.
This article concerns two types of problem related to recovering information about the obstacle $K$ from
certain measurements of scattering rays in the exterior of $K$. These problems have similarities with 
various problems on metric rigidity in Riemannian geometry -- see \cite{kn:SU}, \cite{kn:SUV} and the 
references there for more information.
\subsection{The Scattering Length Spectrum}
The first type of problem deals with the so called scattering length spectrum (SLS). 
Given a scattering ray $\gamma$
in $\Omega_K$, denote by $T_\gamma$ the {\it sojourn time} of $\gamma$ (cf. Sect. 2).  
If $\omega\in \sd$ is the incoming 
direction of $\gamma$ and $\theta\in \sd$ its outgoing direction, $\gamma$  will be called an 
{\it $\ot$-ray}. The {\it scattering length spectrum} of $K$ is defined to be the family 
of sets of real numbers $\sl_K =  \{ \sl_K\ot\}_{\ot}$ where $\ot$ runs over $\ssd$ and
$\sl_K\ot$ is the set of sojourn times $T_\gamma$ of all $\ot$-rays $\gamma$ in $\Omega_K$. 
It is known  (cf. \cite{kn:St3}) that for $d \geq 3$, $d$ odd, and $C^\infty$ boundary $\dk$, we have 
$\sl_K\ot = \mbox{sing supp } s_K(t, \theta, \omega)$
for almost all $\ot$. Here $s_K$ is the {\it scattering kernel} related to the scattering operator for 
the wave equation in $\R\times \Omega_K$ with Dirichlet boundary condition on 
$\R\times \partial \Omega_K$ (cf. \cite{kn:LP1}, \cite{kn:M}). Following \cite{kn:St4}, 
we will say that two obstacles  $K$ and $L$ have {\it almost the same SLS} if there exists a subset 
$\rr$ of full Lebesgue measure in $\ssd$  such that  $\sl_K\ot = \sl_L\ot$ for all $\ot\in \rr$.

It is a natural and rather important problem in inverse scattering by obstacles to get information 
about the obstacle $K$ from its SLS. It is known that various kinds of information about $K$
can be recovered from its SLS, and for some classes of obstacles $K$ is completely recoverable 
(see e.g. \cite{kn:Ma}, \cite{kn:MaR}, \cite{kn:LP2}, \cite{kn:St2}, \cite{kn:St4}) -- for example
star-shaped obstacles are in this class. However, 
as an example of M. Livshits shows (cf. Ch. 5 in \cite{kn:M}; see also Figure 1 on p. 14),  
in general $\sl_K$ does not determine $K$ uniquely. 
\subsection{Travelling Times}
The second type of problems deals with travelling times.
Let $\oo$ be a { large ball} in $\R^d$ containing $K$ in its interior and set $S_0 := \partial \oo$. 
For any pair of points $x,y\in S_0$ consider the  scattering rays $\gamma$  incoming through 
the point $x$ and outgoing through  the point $y$.  Such rays will be called {\it $(x,y)$-geodesics} 
in $\Omega_K$. Given such $\gamma$, let $t_\gamma$ be the {\it length of the part} of $\gamma$ 
from $x$ to $y$. Let $\T_K(x,y)$ the {\it set of travelling times} $t_\gamma$ of all $(x,y)$-geodesic
in $\Omega_K$. If $K$ and $L$ are two obstacles contained in the interior of $S_0$, we will say 
that $K$ and $L$  {\it have almost the same travelling times} if $\T_K(x,y) = \T_L(x,y)$ for almost all 
$(x,y) \in S_0\times S_0$ (with respect to the Lebesgue measure on $S_0\times S_0$).
Our second type of problem is to get information 
about the obstacle $K$ from its travelling times. 

\subsection{Unions of Convex Bodies}
For either kind of problems, we consider obstacles $K$ of the form
\begin{equation}
K = K_1 \cup K_2 \cup \ldots \cup K_{k_0} ,
\end{equation}
where $K_i$ are strictly convex disjoint domains in $\R^d$ ($d\geq 2$) with $C^3$ smooth
boundaries $\dk_i$. In this case the so called generalized Hamiltonian (or bicharacteristic) flow
$\fk_t : S^*(\Omega_K) \longrightarrow S^*(\Omega_K)$ (see Sect. 2) coincides with the billiard flow, so 
it is easier to deal with.

A point $\sigma = (x,\omega)\in S^*(\Omega_K)$ is called {\it non-trapped} if
both curves $\{ \mbox{pr}_1(\fk_t(\sigma)) : t\leq 0\}$ and 
$\{ \mbox{pr}_1(\fk_t(\sigma)) : t\geq 0\}$ in $\Omega_K$ are unbounded. 
Otherwise $\sigma$ is called a {\it trapped point}. Here we use the notation
$\mbox{pr}_1 (y,\eta) = y$ and $\mbox{pr}_2 (y,\eta) = \eta$. Denote by $\trapk$ the {\it set 
of all trapped points}. It is well-known that in general $\trapk$ may have
positive Lebesgue measure and a non-empty interior  in $S^*(\Omega_K)$ (see e.g.
Livshits' example). Set $\dt^*(\Omega_K) = T^*(\Omega_K)\setminus \{ 0\}$.

\bs

\noindent
{\bf Definition 1.1.} Let $K, L$ be two obstacles  in $\R^d$.
We will say that $\Omega_K$ and $\Omega_L$  {\it have conjugate flows} if  there exists a homeomorphism 
$$\Phi : \dt^*(\Omega_{K})\setminus \trapk  \longrightarrow  \dt^*(\Omega_{L})\setminus\trapl$$
which defines a symplectic map on an open dense subset of  $\dt^*(\Omega_{K})\setminus \trapk$,
it maps $S^*(\Omega_K)\setminus \trapk$ onto $S^*(\Omega_L)\setminus \trapl$,
and satisfies $\fl_t\circ \Phi = \Phi\circ \fk_t$ for all $t\in \R$ and $\Phi = \mbox{id}$ on 
$\dt^*(\R^d\setminus \oo)\setminus \trapk = \dt^*(\R^d\setminus \oo)\setminus\trapl$.

\bs

It is known that for $K, L$ in a very large (generic) class of obstacles in $\R^d$ ($d\geq 2$), if $K$ and $L$ have 
almost the same SLS or almost the same travelling times, then $\Omega_K$ and $\Omega_L$ have conjugate 
flows (\cite{kn:St4} and \cite{kn:NS}; see Sect. 2 below where these results are given in full details). 
In this paper we prove 

\bs

\noindent
{\bf Theorem 1.2.} {\it  Let $d \geq 2$ and let each of the obstacles $K$ and $L$ be a 
finite disjoint union of strictly convex domains in $\R^d$ with $C^3$ boundaries. 
If $\Omega_K$ and  $\Omega_L$  have conjugate flows, then $K = L$.}

\bs

For the case where  $\dk$ and $\dl$ are real analytic this was proved in \cite{kn:St2}.
As an immediate consequence of Theorem 1.2 and results in \cite{kn:St4} and \cite{kn:NS},  
one gets the following.

\bs

\noindent
{\bf Corollary 1.3.} {\it Assume that  $K$ and $L$ are obstacles  in $\R^d$ ($d \geq 2$) and each of them is a 
finite disjoint union of strictly convex domains  with $C^3$ boundaries. If $K$ and $L$ have 
almost the same scattering length spectrum, or  $K$ and $L$ have almost the same travelling times, then $K = L$}.

\bs

We remark that the above results are non-trivial. Indeed when $K$ has a large number of
connected components and they are densely packed  (imagine the molecules of a gas in 
a container), then there are a great number of billiard trajectories with large
numbers of reflections (and possibly with many tangencies) in the exterior of $K$. Moreover, 
in such cases many connected components of $K$ can only be reached by billiard trajectories having many 
reflections. So, being able to completely recover the obstacle $K$ by measuring sojourn times only,
or travelling times only, is far from being a trivial matter.
The assumption that $\dk$ and $\dl$ are $C^3$ smooth is required in order to be able to use some
of the results in \cite{kn:St2} and \cite{kn:St4}.

\section{Preliminaries}
\renewcommand{\theequation}{\arabic{section}.\arabic{equation}}

We refer the reader to \cite{kn:MS1}, \cite{kn:MS2} (or Sect. 24.3 in \cite{kn:H}) for   
definition of the generalized Hamiltonian (bicharacteristic) flow on a symplectic manifold with 
boundary. In the case of scattering by an obstacle $K$ this is the {\it generalized geodesic flow} 
$\fk_t : \dt_b^*(\Omega_K) = T_b^*(\Omega_K)\setminus \{ 0\} \longrightarrow
\dt_b^*(\Omega_K)$ generated by the principal symbol of the wave operator in $\R\times \Omega_K$.
Here $T_b^*(\Omega_K) = T^*(\Omega_K)/\sim$ is the  quotient space with respect to the following
equivalence relation on $T^*(\Omega_K)$: $(x,\xi) \sim (y,\eta)$
iff $x = y$ and either $\xi = \eta$ or $x= y \in \dk$ and $\xi$ and $\eta$ are symmetric with respect to the tangent 
plane to $\dk$ at $x$. The image $S_b^*(\Omega_K)$ of the {\it unit cosphere bundle} 
$S^*(\Omega_K)$ under the natural projection is invariant with respect to $\fk_t$. For simplicity
of notation the subscript $b$ will be suppressed, and it will be clear from the context exactly which second
component we have in mind.

In general $\fk_t$ is not a flow in the usual sense of dynamical systems, since there may exist 
different integral curves issued from the same point of the phase space (see \cite{kn:T}).
Let $\kk$ be the { class of obstacles} that  have the following property: for each 
$(x,\xi)\in \dt^* (\partial K)$ if the curvature of $\partial K$ at $x$ vanishes of infinite order 
in direction $\xi$, then all points  $(y,\eta)$ sufficiently close to $(x,\xi)$ are {\it diffractive} (roughly speaking, this means  that $\dk$ is convex at $y$ in the direction of $\eta$). 
It follows that for $K\in \kk$ the flow $\fk_t$ is well-defined and continuous (\cite{kn:MS2}).

We now describe the relevant results from \cite{kn:St4} and \cite{kn:NS} used in the proof
of Corollary 1.3.
Given $\xi\in \sd$ denote by $Z_{\xi}$  
{ the hyperplane in ${\R}^d$ orthogonal to $\xi$ and tangent to $\oo$} such that $\oo$ is 
contained in the open half-space $R_{\xi}$ determined by  $Z_{\xi}$ and having $\xi$ as an inner 
normal. For an $\ot$-ray $\gamma$ in $\Omega$, the {\it sojourn time} $T_{\gamma}$ 
of $\gamma$ is defined by $T_{\gamma} = T'_{\gamma} - 2a$, where $T'_{\gamma}$ is the length of 
that part of $\gamma$ which is contained in $R_{\omega}\cap R_{-\theta}$ and $a$ is the radius of 
the ball $\oo$. It is known (cf. \cite{kn:G}) that this definition does not depend on the choice of 
the ball $\oo$.  
Following \cite{kn:PS2}, given  $\sigma = (x,\xi)\in S^*(\Omega_K)$ so that  
$\gamma_K(\sigma) = \{ \mbox{pr}_1(\fk_t(\sigma)) : t\in \R\}$
is a {\it simply reflecting ray}, i.e. it has no tangencies to $\dk$, we will say that 
$\gamma_K(\sigma)$ is {\it non-degenerate} if for every $t >> 0$ the map
$\R^d \ni y \mapsto \mbox{pr}_2(\fk_t(y,\xi))\in \sd$ is a submersion at $y=x$, i.e. its
differential at $y = x$ has rank $d-1$.
Denote by $\kk_0$ the { class of all obstacles} $K\in\kk$  satisfying the following
non-degeneracy  conditions: $\gamma_K(\sigma)$ is a non-degenerate simply 
reflecting ray for almost all $\sigma\in \sock$ such that $\gk(\sigma)\cap \dk\neq \e$, and $\dk$
does not contain non-trivial open flat subsets ( i.e. open subsets where the curvature is zero at 
every point). It can be shown without much 
difficulty from \cite{kn:PS1} (see Ch. 3 there) that  $\kk_0$ is of second Baire category in 
$\kk$ with respect to the $C^\infty$ Whitney topology in $\kk$. That is, for every $K\in \kk$, 
applying suitable arbitrarily small $C^\infty$ deformations to $\dk$, one gets obstacles from 
the class $\kk_0$ and most deformations have this property.

\bs

\noindent
{\bf Theorem 2.1.} (\cite{kn:St4}) {\it If the obstacles $K, L\in \kk_0$ have almost the same 
SLS,  then $\Omega_K$ and $\Omega_L$ have conjugate flows.
Conversely, if $K, L\in \kk_0$ have conjugate flows, then $K$ and $L$ have the same SLS}.

\bs

There is a similar result for the travelling times spectrum.
Set $S^*_+(S_0) = \{ (x,u) : x\in S_0, u\in \sd \;, \; \la x,u\ra < 0 \}$. 
Consider the cross-sectional map  $\pp_K : S^*_+(S_0)\setminus \trapk \longrightarrow S^*(S_0)$ 
defined by the shift along  the flow $\fk_t$.
 Let $\gamma$ be a  $(x_0,y_0)$-geodesic in $\Omega_K$ for some
$x_0,y_0 \in S_0$, which is a simply reflecting ray. Let $\omega_0\in \sd$ be the 
(incoming) direction of $\gamma$ at $x_0$. 
We will say that $\gamma$ is {\it regular} if the differential of map 
$\sd \ni \omega \mapsto \pp_K (x_0,\omega)\in S_0$ is a submersion at 
$\omega = \omega_0$, i.e. its differential at that point  has rank $d-1$.
Denote by $\ll_0$ the {\it class of all obstacles} $K\in\kk$  such that 
$\dk$ does not contain non-trivial open flat subsets and $\gamma_K(x,u)$ is a regular simply 
reflecting ray for almost all $(x,u) \in S^*_+(S_0)$ such that $\gamma(x,u)\cap \dk\neq \e$. 
Using an argument from Ch. 3 in \cite{kn:PS1} one can show that  $\ll_0$ is of second Baire category 
in $\kk$ with respect to the $C^\infty$ Whitney topology in $\kk$. That is, generic obstacles $K\in \kk$
belong to the class $\ll_0$.

\bs

\noindent
{\bf Theorem 2.2.} (\cite{kn:NS}) {\it If the obstacles $K, L\in \ll_0$ have almost the same travelling times, 
then $\Omega_K$ and $\Omega_L$ have conjugate flows.
Conversely, if $K, L\in \ll_0$ have conjugate flows, then $K$ and $L$ have the same travelling times}.

\bs  

Next, we describe four propositions from \cite{kn:St2} and \cite{kn:St4} that are needed in the proof
of Theorem 1.2. In what follows we assume

\bs

\noindent 
{\bf Hypothesis SCC.} $K$ and $L$ are finite disjoint unions of strictly convex domains
in $\R^d$ ($d\geq 2$) with $C^3$ boundaries, and with  conjugate 
generalized geodesic flows.

\bs 
Given $\sigma\in S^*(\Omega_K)$  denote  $\gk^+(\sigma) := \{ \pr_1(\fk_t(\sigma)) : t\geq 0\}$.

\bs

\noindent
{\bf Proposition 2.3.} (\cite{kn:St2})

(a) {\it There exists a countable family $\{ M_i\} = \{ M_i^{(K)}\}$
of codimension $1$ submanifolds of $S^*_+(S_0)\setminus \trapk$ such that
every $\sigma \in S^*_+(S_0)\setminus (\trapk \cup_i M_i)$ generates
a simply reflecting ray in $\Omega_K$. Moreover the family  $\{ M_i\}$
is locally finite, that is any compact subset of  $S^*_+(S_0)\setminus \trapk$ has common points with
only finitely many of the submanifolds $M_i$.} 

\ms

(b) {\it There exists a countable family $\{ R_i\}$
of codimension $2$ smooth submanifolds of $S^*_+(S_0)$ such that for any
$\sigma\in S^*_+(S_0)\setminus (\cup_i R_i)$ the trajectory $\gk(\sigma)$ has at most one
tangency to $\dk$.}

\ms

(c)  {\it There exists a countable family $\{ Q_i\}$
of codimension $2$ smooth submanifolds of $S^*_{\dk} (\Omega_K)$ such that for any
$\sigma\in S^*_+(S_0)\setminus (\cup_i Q_i)$ the trajectory $\gk(\sigma)$ has at most one
tangency to $\dk$.}

\bs

It follows from the conjugacy of flows and Proposition 4.3 in \cite{kn:St4} that
the submanifolds $M_i$ are the same for $K$ and $L$, i.e. $M_i^{(K)} = M_i^{(L)}$ for all $i$.

\bs 

\noindent
{\bf Remark.} Different submanifolds $M_i$ and $M_j$ may have common points
(these generate rays with more than one tangency to $\dk$) and in general are
not transversal to each other. However, as we see from part (b), if $M_i \neq M_j$
and $\sigma\in M_i \cap M_j$, then locally near $\sigma$, $M_i \neq M_j$, i.e. there exist points in
$M_i\setminus M_j$ arbitrarily close to $\sigma$.

\bs

For the present case we also have some information about the size of the set $\trapk$.

\bs

\noindent
{\bf Proposition 2.4.} (\cite{kn:St2}) {\it Let $d \geq 2$ and let $K$ have the form
{\rm (1.1)}. Then $S^*_+(S_0) \setminus \trapk$ is arc connected.}

\bs

\noindent
{\it Proof.}  This is proved in \cite{kn:St2} assuming $d \geq 3$, however
a small modification of the argument works for $d = 2$ as well. We sketch it here for completeness.
So, assume that $d \geq  2$. Consider an arbitrary 
$\sigma'_0 = (x'_0, \xi'_0)\in S^*_+(S_0)\setminus \trapk$. It is enough
to find a continuous curve $\sigma'(t)$ in $S^*(\R^d\setminus \oo)\setminus \trapk$ such that 
$\sigma'(0) = \sigma'_0$ and $\sigma'(1)$ generates a free trajectory in $S^*(\R^n\setminus \oo)$, 
i.e. a trajectory without any reflections at $\dk$. We will assume that the trajectory $\gk^+(\sigma_0)$ 
has a common point with $K$; otherwise there is nothing to prove.

There exists a strictly convex smooth hypersurface $X$ in $\R^n\setminus \oo$ with 
a continuous unit normal field $\nu_X$ such that for some $x_0 \in X$ and $t > 0$, we have
$x_0' = x_0 + t \nu_X(x_0)$ and $\nu_X(x_0) = \xi'_0$,  and there exists
$x_1\in X$ such that the ray $\{ x_1 + t\nu_X(x_1) : t > 0\}$ has no common points with $K$ 
(e.g.  take the boundary sphere $X$ of a very  large ball in $\R^d\setminus \oo$ with exterior
unit normal field $\nu_X$).
Clearly it is enough to construct a $C^1$ curve $\sigma(t)$ in $S^*(X)\setminus \trapk$ such that 
$\sigma(0) = (x_0,\nu_X(x_0))$ and $\sigma(1) = (x_1, \nu_X(x_1))$.

Set $\tX = \{ (x,\nu_X(x)) : x \in X\}$.
Considering trajectories $\gk^+(\sigma)$ with $\sigma\in \tX$ and using the strict convexity of
$X$ will allow us to use the strong hyperbolicity properties of the billiard flow in $\Omega_K$
(\cite{kn:Si1}, \cite{kn:Si2}; see also \cite{kn:St1}).

Take a very large open ball $U_1$ that contains the orthogonal projection 
of the convex hull $\hat{K}$ of $K$ onto $X$ and the point $x_1$ as well. 
Since $\trapk \cap S^*(U_1)$ is compact and $\sigma_0\notin \trapk$, there exists an open 
connected neighbourhood $V_0$ of $\sigma_0$ in $S^*(U_1)$ with $V_0 \cap \trapk = \e$.

It follows from Proposition 2.3(b) that there exists a countable family
$\{Q'_i\}$ of smooth codimension 2 submanifolds of $S^*(U_1)$ such that for
any $\sigma\in S^*(U_1)\setminus (\cup_i Q'_i)$, the trajectory $\gk(\sigma)$
has at most one tangency to $\dk$. The submanifolds $Q'_i$ are obtained
from the submanifolds $Q_i$ in Proposition 2.3(b) by translation
along the second (vector) component, so, locally they are invariant under the flow $\fk_t$.

Using Thom's Transversality Theorem (cf. e.g. \cite{kn:Hi}), and applying 
an arbitrarily small in the $C^3$ Whitney topology deformation to $X$, as in the proof
of Proposition 5.1 in \cite{kn:St2},  without loss of generality we may assume that
$\tX$ is transversal to each of the submanifolds $Q'_i$. When $d = 2$, this simply means
that $\tX \cap Q'_i$ is a discrete subset of $\tX$. When $d \geq 3$,
$\tX\cap Q'_i$ is a submanifold of $\tX$ with $\dim(\tX\cap Q_i) = (2d-3 + d-1) - (2d-1) = d-3$. 
By the Sum Theorem for the topological dimension $\dim$ (cf. e.g. Theorem 15 in \cite{kn:F}),
for $X' = \tX\cap (\cup_i Q_i)$ we get $\dim(X') \leq d - 3$. 

Next, denote by $X_0$ the set of those $\sigma \in \tX\cap \trapk$ such that
the trajectory $\gk(\sigma)$ has no tangencies to $\dk$.
Given integers $p,q$ such that $q\in \{ 1, \ldots,k_0\}$ and
$p \geq 1$, denote by $X(p,q)$ the set of those
$\sigma \in \tX\cap \trapk$ such that $\gk(\sigma)$ has exactly one
tangency to $\dk$ which is its $p$th reflection point and it
belongs to $\dk_q$.  As in the proof of Proposition 5.1 in \cite{kn:St2} (see 
also the proof of Lemma 3.1 below where we repeat this argument), it
follows that each of the subspaces $X_0$ and $X(p,q)$ of $\tX$ has topological 
dimension $0$.  Assuming $d \geq 3$, the Sum  Theorem for topological 
dimension (cf. Theorem 15 in \cite{kn:F})  shows that 
$$\dim (\tX \cap \trapk) \leq \dim( X' \cup X_0 \cup \cup_{p,q} X(p,q)) \leq d-3 .$$
In the case $d = 2$ we simply have $\dim (\tX \cap \trapk) = 0$.
In both cases a theorem by Mazurkiewicz (see e.g. Theorem 25 in \cite{kn:F}) implies that
$\tX \setminus \trapk$ is arc connected. Thus, there exists a $C^1$ curve $\sigma(t)$ in 
$\tX \setminus \trapk$ (and so in $S^*(X)\setminus \trapk$) such that 
$\sigma(0) = (x_0,\nu_X(x_0))$ and $\sigma(1) = (x_1, \nu_X(x_1))$.
This proves the proposition.
\endofproof

\bs

Since $S^*_+(S_0)$ is a manifold and $\trapk$ is compact, it follows from Proposition 2.4  
that any two points in $S^*_+(S_0)\setminus \trapk$ can be connected by a $C^1$ curve lying entirely 
in $S^*_+(S_0)\setminus \trapk$.
Denote by $S_K$ the { set of the points $\sigma\in S^*_+(S_0)\setminus \trapk$
such that $\gamma_K(\sigma)$ is a simply reflecting ray}. It follows from \cite{kn:MS2} (cf. also
Sect. 24.3 in \cite{kn:H}) and Proposition 2.4 in \cite{kn:St2}) that $S_K$ is open and dense
and has full Lebesgue measure in $\sock$. 
Proposition 4.3 in \cite{kn:St4} shows that if $K,L$ have conjugate flows, then $S_K = S_L$.
Moreover, using Proposition 6.3 in  \cite{kn:St4} and the above Proposition 2.4 we get the following.
 
\bs

\noindent
{\bf Proposition 2.5.} {\it Let $K, L$ satisfy Hypothesis SCC. Then
\begin{equation}
\#(\gamma_K(\sigma) \cap \dk) = \# (\gamma_L(\sigma)\cap \dl) 
\end{equation}
for all $\sigma\in S_K = S_L$.}

\bs

\noindent
{\bf Definition.} A $C^1$ path $\sigma(s)$, $0\leq s\leq a$ (for some $a > 0$), in $\sock$ will be called 
$K$-{\it admissible} if it has the following properties:

\begin{enumerate}

\item[(a)] $\sigma(0)$ generates a {\it free ray} in $\Omega_K$,
i.e. a  ray without  any common points with $\dk$.

\item[(b)]  if  $\sigma(s)\in M_i$ for some $i$ and $s\in [0,a]$, then $\sigma$ is
transversal to $M_i$ at $\sigma(s)$ and $\sigma(s) \notin M_j$ for any 
submanifold $M_j\neq M_i$ .

\end{enumerate}

\ms

Notice that under Hypothesis SCC, we have $\sock = \socl$ and a path is
$K$-admissible iff it is $L$-admissible. So, in what follows we will just call such paths 
{\it admissible}.
If a path $\sigma$ is admissible, it follows from (b) that every $\sigma(s)$ generates a scattering ray 
with at most one tangent point to $\dk$ and the tangency (if any) is of first order only.
It is clear that if the curve $\sigma(s)$ ($0\leq s\leq a$) is admissible and $\rho(s)$ 
($0\leq s\leq a$) is  uniformly close  to $\sigma(s)$ (i.e. $\rho(s)$ and $\sigma(s)$ and their derivatives
are $\ep$-close for all $s$ for some sufficiently small $\ep > 0$), then $\rho(s)$ is also admissible.

\bs

\noindent
{\bf Proposition 2.6.} {\it For any $\rho\in \sock$ there exists an admissible path  
$\sigma(s)$, $0 \leq s\leq a$, with $\sigma(a) = \rho$.}

\bs

\noindent
{\it Proof.} This follows from Proposition 6.3 in \cite{kn:St4} (or rather its proof) and Proposition 2.4 above.
\endofproof

\section{Proof of Theorem 1.2}
\renewcommand{\theequation}{\arabic{section}.\arabic{equation}}
 
Let $K$ and $L$ be as in Theorem 1.2. We will show that they coincide.
A point $y \in \dk$ will be called {\it regular} if $\dk = \dl$ in an open neighbourhood
of $y$ in $\dk$. Otherwise $y$ will be called irregular.
A point $y \in \dk$ will be called {\it accessible} if there exists $\rho \in S^*_+(S_0)\setminus \trapk$ 
such that the trajectory $\gk(\rho)$ contains the point $y$. Denote by $\aa_K$ { the set
of all accessible points} $y\in \dk$.

\bs

\noindent
{\bf Lemma 3.1.} {\it  Let $d \geq 2$ and let $K$ have the form {\rm (1.1)}. Then $\aa_K$ is dense in $\dk$.}

\bs

\noindent
{\it Proof.} We will use a slight modification of an argument contained in the proof of Proposition
5.1 in \cite{kn:St2} (see also Proposition 2.4 above) which assumes $d \geq 3$. We will sketch this proof here
dealing with the case $d = 2$ as well. 

First, it follows from Proposition 2.3(c) that there exists a countable locally finite
family $\{ Q_i\}$ of codimension $2$ smooth submanifolds of $S^*_{\dk} (\Omega_K)$ such that for any
$\sigma\in S^*_{\dk} (\Omega_K) \setminus (\cup_i Q_i)$ the trajectory $\gk(\sigma)$ has at most one
tangency to $\dk$. Notice that when $d = 2$, we have $\dim(S^*_{\dk} (\Omega_K)) = 2$, and then
each $Q_i$ is a finite subset of $S^*_{\dk} (\Omega_K)$, so  $\cup_i Q_i$
is a discrete subset of $S^*_{\dk} (\Omega_K)$.

Consider an arbitrary $x_0 \in \dk$ and let $\delta > 0$. We will show that there exist points $x\in \aa_K$ 
which is $\delta$-close to $x_0$. Take an arbitrary $u_0\in \sd$ so that 
$\sigma_0 = (x_0,u_0) \in S^*_{\dk} (\Omega_K)$.  Fix a small $\ep > 0$ and set 
$X = \{ x_0 + \ep \, u : \|u-u_0\| < \ep \}$ and  $\tX = \{ (x_0  + \ep \, u, u) : \|u-u_0\| \}$. 
As in the proof of Proposition 2.4, using Thom's Transversality 
Theorem (cf. e.g. \cite{kn:Hi}), and applying  an arbitrarily small in the $C^3$ Whitney topology deformation 
to $X$,  we get a $C^3$ convex surface $Y$ in  $\Omega_K$ which $C^3$-close to $X$ and so that
$\tY$ is transversal to each of the submanifolds $Q'_i$. We take $Y$ so close to $X$ that for every 
$(y,v) \in \tY$ for the point $x\in \dk$ with $x+ t \, v = y$ for some $t> 0$ close to $\ep$ we have
$\|x_0-x\| < \delta$.
As in the proof of Proposition 2.4, when $d = 2$, $\tY \cap Q'_i$ is a discrete subset of $\tY$, and
when $d \geq 3$,  for $Y' = \tY\cap (\cup_i Q_i)$ we have $\dim(Y') \leq d - 3$.

Let $Y_0$ the set of those $\sigma \in \tY \cap \trapk$ such that
the trajectory $\gk(\sigma)$ has no tangencies to $\dk$, and let
$\tF = \prod_{r=0}^\infty F$, where $F = \{ 1, 2, \ldots,k_0\}$.
Fix an arbitrary $\theta \in (0,1)$ and consider the metric $d$ on $\tF$ defined
by $d( (x_i), (y_i)) = 0$ if $x_i = y_i$ for all $i$,  $d( (x_i), (y_i)) = 1$ if $x_0 \neq y_0$, and
 $d( (x_i), (y_i)) = \theta^N$ if $x_i = y_i$ for all $i = 0,1 \ldots, N-1$ and $N \geq 1$ is maximal
with this property. Then $\tF$ is a compact totally disconnected metric space.
Consider the map $f : Y_0 \longrightarrow \tF$, defined by
$f(\sigma) = (i_0, i_1, \ldots)$, where the $j$th reflection point
of $\gk(\sigma)$ belongs to $\dk_{i_j}$ for all $j = 0,1, \ldots$.
Clearly, the map $f$ is continuous and, using the strict convexity of $Y$, it follows from \cite{kn:St1}
that $f^{-1} : f(Y_0) \longrightarrow Y_0$ is also continuous (it is
in fact Lipschitz with respect to an appropriate metric on $\tF$).
Thus, $Y_0$ is homeomorphic to a subspace of $\tF$ and therefore
$Y_0$ is a compact totally disconnected subset of $\tY$.

Given integers $p,q$ such that $q\in \{ 1, \ldots,k_0\}$ and
$p \geq 1$, denote by $Y(p,q)$ the set of those
$\sigma \in \tY\cap \trapk$ such that $\gk(\sigma)$ has exactly one
tangency to $\dk$ which is its $p$th reflection point and it
belongs to $\dk_q$. Given $p, q$, define
$f : Y(p,q) \longrightarrow \tF$ as above and notice that in
the present case we have $i_q = p$ for any $\sigma\in Y(p,q)$,
where $f(\sigma) = (i_j)$. This and the definition of $Y(p,q)$
imply that $f$ is continuous on $Y(p,q)$ and using \cite{kn:St1}
again, we get that $Y(p,q)$ is homeomorphic to a subspace of $\tF$
and so $Y(p,q)$ is a compact totally disconnected subset of  $S^*_{\dk} (\Omega_K)$. 

Thus, each of the sets $Y'$, $Y_0$ and $Y(p,q)$ is a subsets of first
Baire category in  $\tY$. It now follows from  Baire's category theorem that
$Y' \cup Y_0 \cup \cup_{p,q} Y(p,q)$ is nowhere dense in $\tY$.
Thus, $\aa_K \cap \tY \supseteq \tY \setminus (Y' \cup Y_0 \cup \cup_{p,q} Y(p,q))$
is dense in $\tY$. 

Thus, there exists $(y,v) \in \tY$ which generates a non-trapped trajectory. By the choice of
$Y$, if $x\in \dk$ is so that $x+ t \, v = y$ for some $t> 0$ close to $\ep$, then we have
$\|x_0-x\| < \delta$. Clearly, $x\in \aa_K$, so $x_0$ is $\delta$-close to a point in $\aa_K$.
This proves the assertion.
\endofproof

\bs

\noindent
{\bf Definition.} 
For any integer $n \geq 1$ let $Z_n$ be { the set of those irregular points} $x\in \dk$ for which
there exists an admissible path $\sigma(s)$, $0\leq s\leq a$, in $S^*_+(S_0)\setminus \trapk$ such that
$\sigma(a)$ generates a free ray in $\R^d$, $x$ belongs to the billiard trajectory
$\gk^+(\sigma(a))$ and for any $s\in [0,a]$ the trajectory $\gk^+(\sigma(s))$ has at
most $n$ irregular common points with $\dk$.

\bs

Notice that in the above definition, the billiard trajectory $\gk^+(\sigma(s))$ may have more than $n$
common points with $\dk$ -- at most $n$ of them will be irregular and all the others must be regular.
We will prove by induction on $n$ that $Z_n = \e$ for all $n \geq 1$.

First, we will show that $Z_1 = \e$. 
Let $x\in Z_1$. Then there exists a $C^1$ path $\sigma(s)$, $0\leq s\leq a$, in $S^*_+(S_0)$ 
as in the definition of $Z_{1}$. In particular, $x$ lies on $\gk^+(\sigma(a))$ and for each 
$s\in [0,a]$ the trajectory $\gk^+(\sigma (s))$ has at most $1$ irregular point. As in the argument 
below dealing with the inductive step, we may assume that $a > 0$ is the smallest number for which 
$\gk^+(\sigma (a))$ has an irregular point, i.e. for all $s\in [0,a)$ the trajectory
$\gk^+(\sigma (s))$ contains no irregular points\footnote{We should probably remark that 
$\gamma$ may have more than one common points with $\dk$,
and for some $s < a$, $\gk^+(\sigma (s))$ may have common points with $\dk$, however all of them
will be regular points.} at all. Set $\rho = \sigma(a)$ and $\gamma = \gk^+(\sigma(a))$.

Before continuing let us remark that $\gamma = \gk^+(\rho)$ and $\gamma' = \gl^+(\rho)$ must have 
the same number of common points with $\dk$ and $\dl$, respectively. Indeed, if $\gamma$ has a tangent 
point to $\dk$ (then $\rho \in M_i$ for some, unique, $i$), then $\gamma$ has only one tangent point 
to $\dk$ and $\gamma'$ also has a unique tangent point to $\dl$. Let $k$ be the number of proper reflection 
points of $\gamma$ at $\dk$. Then we can find $\rho' \in S^*_+(S_0)\setminus \trapk$ arbitrarily close to 
$\rho$ so that $\gk^+(\rho')$ has exactly $k$ common points with $\dk$, all of them being proper reflection 
points. By Proposition 2.5, $\gl^+(\rho')$ also has exactly $k$ common points with $\dl$, all of them being 
proper reflection points. Thus, $\gl^+(\rho')$ also has at most $k$ proper reflection points, so
$$\sharp \mbox{\rm (proper reflection points of $\gk^+(\rho)) 
\geq \sharp$ (proper reflection points of }  \gl^+(\rho)) .$$ 
By symmetry,  it follows that $\gk^+(\rho)$ and $\gl^+(\rho)$  have the same number of proper reflection points,
and therefore the same number of common points with $\dk$ and $\dl$, respectively.

Let $x_1, \ldots, x_m$ be the 
common points of $\gk(\rho)$ and $\dk$ and let $x_i = x$ for some $i$. Then for each $j\neq i$
there exists an open subset $U_j$ of $\dk$ such that $U_j = U_j \cap \dl$. Let $\rho = (x_0,u_0)$
and let $\omega = (x_{m+1}, u_{m+1}) \in S^*(S_0)$ be the last common point of $\gk(\rho)$
with $S^*(S_0)$ before it goes to $\infty$. Let $\fk_{t_j}(\rho) = (x_j,u_j)$, $1 \leq j \leq m+1$.
It then follows from the above that
$\fk_t(\rho) = \fl_t(\rho)$ for $0 \leq t < t_i$ and also, $\fk_{\tau}(\omega) = \fl_{\tau}(\omega)$
for all $- (t_{m+1} - t_i) < \tau \leq 0$. So, the trajectories $\gk(\rho)$ and $\gl(\rho)$ both pass
through $x_{i-1}$ with the same (reflected) direction $u_{i-1}$ and through $x_{i+1}$ with the same
(reflected) direction $u_{i+1}$. Thus, $x_1, \ldots, x_{i-1}, x_{i+1}, \ldots, x_m$ are common points of
$\gl^+(\rho)$ and $\dl$. As observed above,  $\gl^+(\rho)$ must have exactly $m$ common points 
with $\dl$, so it has a common point $y_i$ with $\dl$ `between' $x_{i-1}$ and $x_{i+1}$.

Next, we consider two cases.

\ms

\noindent
{\bf Case 1.} $x_i$ is a proper reflection point of $\gamma$ at $\dk$. It then follows immediately from 
the above that $x_i$ lies on $\gamma'$ and $\gamma'$ has a proper reflection point at $x_i$, so in
particular $x_i \in \dl$. Moreover for any $y\in \dk$ sufficiently close to $x_i$ there exists 
$\rho'\in S^*_+(S_0)\setminus \trapk$ close to $\rho$ so that $\gk^+(\rho')$ has a proper reflection point at
$y$. Then  repeating the above argument and using again $U_j = U_j \cap \dl$ for $j \neq i$, shows that 
$y \in \dl$. Thus, $\dk = \dl$ in an open neighbourhood of $x = x_i$ in $\dk$, which is a contradiction
with the assumption that $x$ is an irregular point.

\ms

\noindent
{\bf Case 2.} $x_i$ is a tangent point of $\gamma$ to $\dk$. Then $\gamma$ (and so $\gamma'$) has 
$m-1$ proper reflection points and  $y_i$ is a point on the segment $[x_{i-1}, x_{i+1}]$.
Assume for a moment that $y_i \neq x_i$. Clearly we can choose $x'_i \in \dk$ arbitrarily close to $x_i$ and 
$u'_i\in \sd$ close to $u_i$ so that $u'_i$ is tangent to $\dk$ at $x'_i$ and the straight line determined by $x'_i$
and $u'_i$ intersects $\dl$ transversally near $y_i$. Let $\rho'\in S^*_+(S_0)\setminus \trapk$ be the point close
to $\rho$ which determines a trajectory $\gk^+(\rho')$ passing through $x'_i$ in direction $u'_i$, i.e. tangent
to $\dk$ at $x'_i$. Then $\gk^+(\rho')$ has $m-1$ proper reflection points and one tangent point, while 
$\gl^+(\rho')$ has $m$ proper reflection points and no tangent points at all. This impossible, so we must have
$y_i = x_i$. Moreover, using a similar argument one shows that every $x' \in \dk$ sufficiently close to $x_i$
belongs to $\dl$, as well. So, $x_i$ is a regular point, contradicting our assumption. 

\bs 

\noindent 
This proves that $Z_1 = \e$.
Next, suppose that for some $n > 1$ we have $Z_1 = \ldots = Z_{n-1} = \e$ and  that $Z_n \neq \e$. Let $x\in Z_{n}$.
Then there exists an admissible  $C^1$ path $\sigma(s)$, $0\leq s\leq a$, in $S^*_+(S_0)$ 
as in the definition of $Z_{n}$.
In particular, $x$ lies on $\gk^+(\sigma(a))$ and for each $s\in [0,a]$ the trajectory 
$\gk(\sigma (s))$ has at most $n$ irregular points. Set
$$A = \{s \in (0,a] : \gk^+(\sigma(s)) \; \mbox{\rm has $n$ irregular points }\, \} .$$
Clearly $A \neq \e$, since by assumption $x\in Z_n$ and $Z_j = \e$ for $j < n$.
Set $b = \inf A$. Then there exists a decreasing sequence $\{s_m\} \subset (0,a]$ with
$s_m \searrow b$ and such that for each $m \geq 1$ the trajectory $\gk^+(\sigma(s_m))$
contains $n$ distinct irregular (consecutive) points $x_1^{(m)}, \ldots, x_n^{(m)}$.
Since $K$ has the form (1.1), these points belong to distinct connected components of $K$.
Choosing an appropriate subsequence of $\{ s_m\}$, we may assume that $x_i^{(m)}$ belongs
to the same connected component $K_{j_i}$ for all $m\geq 1$ and there exists
$x_i = \lim_{m\to \infty} x_i^{(m)}$ for all $i = 1, \ldots,n$. Then $x_1, \ldots, x_n$
are distinct common points of $\gk^+(\sigma(b))$ with $\dk$. Moreover, if some $x_i$
is a regular point, then $\dk = \dl$ in an open neighbourhood of $x_i$ in $\dk$, so
$x_i^{(m)}$ would be regular for  large $m$ -- contradiction. Thus, all $x_1, \ldots, x_n$
are irregular points. 

The above reasoning shows that we may assume $b = a$,  $\gk^+((a))$ contains exactly
$n$ irregular points, and for  any $s\in [0,a)$ the trajectory $\gk^+(\sigma(s))$ has
$< n$ irregular points. Now the inductive assumption implies that { for any $s\in [0,a)$ the 
trajectory $\gk^+(\sigma(s))$ contains no irregular points at all}. Since $Z_j = \e$ for $j < n$, 
every irregular point of $\gk^+(\sigma(a))$ must belong to $Z_n$. 

Set $\gamma = \gk^+(\sigma(a))$ for brevity.
Let $x_1, \ldots, x_n$ be the consecutive irregular points of $\gamma$ (which may have some 
other common points with $\dk$) and for
$s < a$ close to $a$, let $x_1(s), \ldots, x_n(s)$ be the consecutive common points of 
$\gk^+(\sigma(s))$ with $\dk$ such that $x_i(s)$ lies on the connected component $K_{j_i}$ of 
$K$ containing $x_i$ ($i = 1, \ldots,n$). Then for every $i$, $\dk = \dl$ in an open neighbourhood of $x_i(s)$
in $\dk$ for $s< a$ close to $a$, so there exists an open subset $U_i$ of $\dk$ with
$x_i \in \overline{U_i}$ and $\dk \cap U_i = \dl \cap U_i$.

\ms 
\noindent 
Next, we consider two cases.

\ms

\noindent
{\bf Case 1.}  $\gamma$ contains no tangent points to $\dk$.  

\noindent 
Choose a small $\delta > 0$ (how small will be determined later). We will replace the path
$\sigma(s)$ by another one $\tsigma(s)$, $0 \leq s \leq a$, such that $\tsigma(s) = \sigma(s)$ for
$s\in [0,a-2\delta]$.

\ms 
 \noindent 
Let $\fk_{t_1} (x_1,u_1) = (x_2,u_2)$ for some $t_1 > 0$, 
where $u_1\in \sd$ is the (reflected) direction of $\gamma$ at $x_1$ and
$u_2\in \sd$ is the reflected direction of $\gamma$ at $x_2$.
Take a small $\epsilon > 0$ and set 
$$X := \{ x_1 + \ep \, u : u\in \sd \; , \, \|u-u_1\| < \delta  \} \quad, \quad 
\tX := \{ (x_1 + \ep \, u, u)  : u\in \sd \; , \, \|u-u_1\| < \delta  \} .$$

\noindent 
Let $u_1(s)$ be the (reflected) direction of the trajectory $\sigma (s)$ at $x_1(s)$.
Take $t' < t_1$ close to $t_1$ and set  $\tY = \fk_{t'}(\tX)$, $Y = \pr_1(\tY)$.  Then
$$\tY = \{ (y,\nu_Y(y)) : y \in Y\} ,$$
where $\nu_Y(y)$ is the {\it unit normal} to $Y$ at $y$ in the direction of the flow $\fk_t$, and
$(y_1,v_1) = \fk_{t'} (x_1,u_1)\in \tY$, so $y_1 \in Y$.

Take a small $\ep > 0$ and consider 
$$\hY = \{ (y,v) : y \in Y, v\in \sd , \|v- \nu_Y(y)\| < \ep\} .$$
There exists an open neighbourhood $V$ of $(x_1,u_1)$  in  $S^*_{\dk}(\Omega_K)$ such that
the shift\\ $\Phi: V \longrightarrow \hY$  along the flow $\fk_t$ is well-defined and smooth. 
Assuming $\delta$ is sufficiently small,
$(y_1(s), v_1(s)) = \Phi(x_1(s), u_1(s))$ is well-defined for $s\in [a-2\delta, a]$. Then
$y_1(s)$, $s\in [a-2\delta, a]$, is a $C^1$ curve on 
$Y$ and the ray issued from $y_1(s)$ in direction $v_1(s)$ hits $\dk_{j_2}$ at $x_2(s)$.
Moreover, $v_1(a) = \nu_Y(y_1)$.

Assuming $\delta > 0$ is sufficiently small,
for all $s,s'\in [a-2\delta, a]$, there exists a unique vector $v_1(s,s')\in \sd$ such that
$$y_1(s) + t(s,s') v_1(s,s') = x_2(s')$$
for some $t(s,s')$ close to $t_1-t'$. Moreover, $v_1(s,s')$ and $t(s,s')$ are smoothly ($C^1$)
depending on $s$ and $s'$. Clearly, $v_1(s,s) = v_1(s)$.

Since $Y$ is a strictly convex surface with a unit normal field $\nu_Y(y)$, 
it is clear that for any $s\in (0,a]$ sufficiently close to $a$ there exists $y\in Y$ close
to $y_1$ such that $y + t\nu_Y(y) = x_2(s)$ for some $t$ close to $t_1-t'$. Fix a small $\delta > 0$,
set $s_0 = a -\delta/2$ and let $\ty\in Y$ be so that 
\begin{equation}
\ty + t \, \nu_Y(\ty) = x_2(s_0) \in U_2 .
\end{equation}
Take a $C^1$ curve $\ty_1(s)$, $s\in [a-2\delta, a]$, such that 
$\ty_1(s) = y_1(s)$ for $s\in [a-2\delta, a- 3\delta/2]$ and $\ty_1(a) = \ty$.

Next, define $\hs(s)$, $s\in [0,a]$, by $\hs(s) = s$ for $0 \leq s\leq a-\delta$ and
$$\hs(s) = \frac{s}{2} + s_0 - \frac{a}{2} = s_0 - \frac{a-s}{2} \in [a-\delta ,s_0]
\quad ,\quad s\in [a-\delta,a] .$$
Then $\hs$ is continuous however not differentiable at $s = a-\delta$. 
Take a $C^1$ function $\ts(s)$, $s\in [0,a]$, which coincides with $\hs$ on $[0,a-\delta - \delta_1]$
and on $[a-\delta+\delta_1,a]$ for some $\delta_1  < \delta/2$ so that the range of $\ts$ is the same
as that of $\hs$, i.e. it coincides with the interval $[0,s_0]$. Now for any $s\in [a-2\delta, a]$ take
$\tv_1 (s)$ so that 
$$\ty_1(s) + \tt(s) \tv_1(s) = x_2(\ts(s)) \quad , \quad s\in [a-2\delta, a] ,$$
for some $\tt(s)$ close to $t_1-t'$. For $s\in [a-2\delta, a-\delta - \delta_1]$ we have
$\ty_1(s) = y_1(s)$ and $\ts(s) = s$, which imply $\tv_1(s) = v_1(s)$.
When $s= a$ we have $\ty_1(a) =\ty$ and $\ts(a) = \hs(a) = s_0$, so by (3.3) we must have
$\tv_1(a) = \nu_Y(\ty)$, i.e. $(\ty_1(a), \tv_1(a)) \in \tY$, and therefore 
\begin{equation}
\pr_1(\Phi^{-1} (\ty_1(a), \tv_1(a))) = \pr_1(\fk_{-t'} (\ty_1(a), \tv_1(a))) = x_1 .
\end{equation}


Set $(\tx_1(s), \tu_1(s)) = \Phi^{-1}(\ty_1(s), \tv_1(s))$, $s \in [a-2\delta, a]$. For
$s\in [a-2\delta, a-\delta-\delta_1]$, we have $\ts(s) = \hs(s) = s$ and therefore
$\ty_1(s) = y_1(s)$ and $\tv_1(s) = v_1(s)$, which gives $(\tx_1(s), \tu_1(s)) = (x_1(s),u_1(s))$.

Now define the path $\tsigma(s)$, $0 \leq s\leq a$, on $S^*_+(S_0)$ by
$\tsigma(s) = \sigma(s)$ for $s\in [0, a-2\delta]$ and for $s\in [a-2\delta, a]$ let $\tsigma(s)$
be the unique point such that $\fk_{\tt_0(s)}(\tsigma(s)) = (\tx_1(s), \tu_1(s))$ for some
$\tt_0(s)$ close to $t_0(s)$, where $\fk_{t_0(s)}(\sigma(s)) = (x_1(s), u_1(s))$. It follows from (3.4) that
$\tsigma(a)$ contains the point $x_1$. Moreover the construction of $\tsigma$ shows that it is a $C^1$ path.
Assuming $\delta$ is sufficiently small, $\tsigma(s)$ has no tangent points to $\dk$ for all
$s\in [a-2\delta, a]$, so $\tsigma(s)$ is an admissible path. For $s\in [0,a-2\delta]$, 
$\gk(\tsigma(s)) = \gk(\sigma(s))$ contains no irregular points. For $s\in [a-2\delta,a]$,
$\gk(\tsigma(s))$ has at most $n$ irregular points, close to $x_1, x_2, \ldots,x_n$ (if any).
If for some $s\in [a-2\delta,a]$, $\gk(\tsigma(s))$ has exactly $n$ irregular points, then 
the second of these must be $x_2(\ts(s))$. However, $\ts(s) < a$ for all such $s$, so 
$x_2(\ts(s)) \in U_2$, and therefore that cannot be the case. Thus, for all $s\in [a-2\delta,a]$, 
$\gk(\tsigma(s))$ has  not more than $n-1$ irregular points, and therefore $x_1$ (and every 
other irregular  point on $\gk(\tsigma(a))$) belongs to $Z_{n-1}$. This is a contradiction with 
the inductive assumption that $Z_j = \e$ for all $j = 1,\ldots,n-1$.

\ms

\noindent
{\bf Case 2.} $\gamma$ contains a tangent point $y_0$ to $\dk$ (this may be one of the irregular points $x_i$).

\ms
\noindent 
Let $y_0 \in \dk_p$ for some $p$. 
According to the definition of an admissible path, $\gamma$ has only one tangent point to $\dk$, so all other 
common points are proper reflection points.   Since $n > 1$, at least one of the irregular points on $\gamma$ 
is a proper reflection point, and at least one of them is different from $y_0$. We will assume $x_1 \neq y_0$;
the general case is very similar.
As in Case 1, we will replace the path $\sigma(s)$ by another one $\tsigma(s)$, $0 \leq s \leq a$, such that 
$\tsigma(s) = \sigma(s)$ for $s\in [0,a-2\delta]$ for some small $\delta > 0$.

Let $\fk_{\tau} (x_{1},u_1) = (y_0,v_0)$ for some $\tau \in \R$ (which may be positive or negative), 
where $u_1\in \sd$ is the (reflected) direction of $\gamma$ at $x_1$ and 
$v_0$ is the  direction of $\gamma$ at $y_0$. Take a small $\epsilon > 0$ and let 
$$X := \{ x_1 + \ep \, u : u\in \sd \; , \, \|u-u_1\| < \delta  \} \quad, \quad 
\tX := \{ (x_1 + \ep \, u, u)  : u\in \sd \; , \, \|u-u_1\| < \delta  \} .$$
Assuming $\delta > 0$ is small enough, for $s\in [a-2\delta, a]$ the trajectory $\gk(\sigma(s))$ has a 
proper reflection point $x_1(s) \in \dk_{j_1}$.
Let $u_1(s)$ be the (reflected) direction of the trajectory $\sigma (s)$ at $x_1(s)$.

Clearly in the present case we have $\sigma(a) \in M_i$ for some $i$. By the definition of an admissible
path, $\sigma(a) \notin M_j$ for any $j\neq i$ and moreover $\sigma$ is transversal to $M_i$ at $\sigma(a)$.
Let $\sigma(s) = (x_0(s), u_0(s)) \in S^*_+(S_0)$ and let $\fk_{t_0(s)} (x_0(s),u_0(s)) =  (x_1(s),u_1(s))$.
The shift $\Phi:  S^*_+(S_0) \longrightarrow S^*_{\dk}(\Omega_K)$ 
along the flow $\fk_t$ is well-defined on an open neighbourhood $V_0$ of  $(x_0,u_0) = (x_0(a), u_0(a))$ and 
defines a  diffeomorphism $\Phi:  V_0 \longrightarrow V = \Phi(V_0)$ for some small open neighbourhood
$V$ of $(x_1,u_1)$ in $S^*_{\dk}(\Omega_K)$. Since $M_i$ is a submanifold of $S^*_+(S_0)$
of codimension  one, we can take $V_0$ so that $V_0\setminus M_i$ has two (open) connected components 
(separated  by $M_i$) -- each of them diffeomorphic to an open half-ball. 
We take $V_0$ so small that $V_0 \cap M_j = \e$ for all $j \neq i$.

Setting $M'_i = \Phi(M_i \cap V_0)$,  we get a similar picture in $V$, 
namely $V\setminus M'_i$ has two (open) connected components 
(separated  by $M'_i$) -- each of them diffeomorphic to an open half-ball.
Assuming $\delta$ is sufficiently small, we have $\sigma(s) \in V_0$ for all 
$s\in [a-2\delta,a]$ and $\sigma$ is transversal to $M_i$ at $\sigma(a) = (x_0,u_0)$. 
Thus, the curve $(x_1(s),u_1(s))$, 
$s\in [a-2\delta,a]$, in $V$ is transversal to $M'_i$ at $(x_1,u_1)$, 
so it must be contained in one of the connected 
components of $V \setminus M'_i$. Denote by $V_+$ the connected component of $V \setminus M'_i$ that contains 
$(x_1(s), u_1(s))$ for $s\in [a- 2\delta, a)$.

Finally, to get this picture near the tangent point $y_0$, take $\tau' \in (\tau-\delta, \tau- \delta/2)$ and set
$$F = \{ (x_1, u) : u\in \sd\; , \|u-u_1\| < \delta\} \subset V \quad, \quad \tY = \fk_{\tau'} (F) ,$$
and   $Y = \pr_1(\tY)$.  Then $\tY = \{ (y,\nu_Y(y)) : y \in Y\}$,
where $\nu_Y(y)$ is the {\it unit normal} to $Y$ at $y$ in the direction of the flow $\fk_t$, and
$(y_1,v_1) = \fk_{\tau'} (x_1,u_1)\in \tY$, so $y_1 \in Y$. Let $\Psi : F \longrightarrow \tY$ 
be the shift along the
flow $\fk_t$.  Assuming that $\delta$ is sufficiently small, this defines a diffeomorphism 
$\Psi : F \longrightarrow G$ between $F$ and an open subset $G$ of $\tY$. Set $M''_i = \Psi(F\cap M'_i)$.
Then $M''_i$ consist of those $(y,\nu_Y(y)) \in G$ that generate trajectories tangent to $\dk$ 
(and this can only happen in the vicinity of $y_0$ on $\dk_p$). It is clear  (by a direct observation 
using the  convexity of $K_{p}$) that $G \setminus M''_i$ has two connected components.  Let $G_+$ 
be the one with $\Psi^{-1} (G_+) \subset V_+$. Thus, there exists $v'_1 \in \sd$, $\|v'_1 - v_1\| < \delta$ 
such that $(y_1,v'_1) \in G_+$.  Applying $\Psi^{-1}$, this gives
$u'_1 \in \sd$ with $\|u'_1 - u_1\| < \delta$ such that $(x_1,u'_1) \in V_+$.  

Since $V_+$ is connected (in fact diffeomorphic to an open half-ball), there exists a 
$C^1$ curve $(\tx_1(s), \tu_1(s))$,  $s\in [a-2\delta, a]$, in $V_+$ such that 
$(\tx_1(s), \tu_1(s)) = (x_1(s), u_1(s))$ for $s\in [a-2\delta, a-\delta]$ and $(\tx_1(a), \tu_1(a)) = (x_1,u'_1)$.
Define the path $\tsigma(s)$ by  $\tsigma(s) = \sigma(s)$ for $s\in [0,a-2\delta]$
and $\tsigma(s) = \Phi^{-1}(\tx_1(s), \tu_1(s))$ for  $s\in [a-2\delta, a]$. It is clear from the 
construction that $\tsigma$ is a $C^1$ path in $S^*_+(S_0)$. From the properties of $\sigma$, we have that 
for $s\in [0,a-2\delta]$ the trajectory $\gk(\tsigma(s))$ does not contain any irregular points. The 
choice $V_0$ and that of the curve $(\tx_i(s), \tu_1(s))$, $s\in [a-2\delta,a]$, show that $\gk(\tsigma(s))$ 
has no tangencies to $\dk$ for any $s\in [a-2\delta,a]$. On the other hand, $\gk(\tsigma(a))$ contains the 
irregular point $x_1$ (and therefore must have $n$ irregular points, since $Z_j = \e$ for $j < n$). Now 
repeating the argument from Case 1 we get a contradiction.
Thus, we must have $Z_n = \e$.

\ms 
\noindent 
This completes the induction process and proves that $Z_n = \e$ for all $n \geq 1$. 

We will now prove that $\dk \subseteq \dl$. By Lemma 3.1, $\aa_K$ is dense in $\dk$, so
it is enough to show that $\aa_K \subseteq \dl$. Given $x\in \aa_K$, there exists 
$\rho \in S^*_+(S_0) \setminus \trapk$ such that $\gk^+(\rho)$ has a reflection point at $x$. Then 
there exists an admissible path $\sigma(s)$, $0\leq s \leq a$, for some $a > 0$, 
with $\rho' = \sigma(a)$ arbitrarily close to $\rho$. Then $\gk^+(\sigma(\rho'))$ has a reflection
point $x'$ at $\dk$ near $x$. Since there are no irregular points on
$\sigma(a)$, it follows that $\dk = \dl$ on an open neighbourhood of $x'$. Thus, $x$ is arbitrarily close
to $\dl$, so we must have $x\in \dl$. This proves that $\dk \subseteq \dl$. 

By symmetry, $\dl\subseteq \dk$, so we have $\dk = \dl$.
\endofproof

\bs


\bs

\clearpage
\includegraphics[scale=0.6]{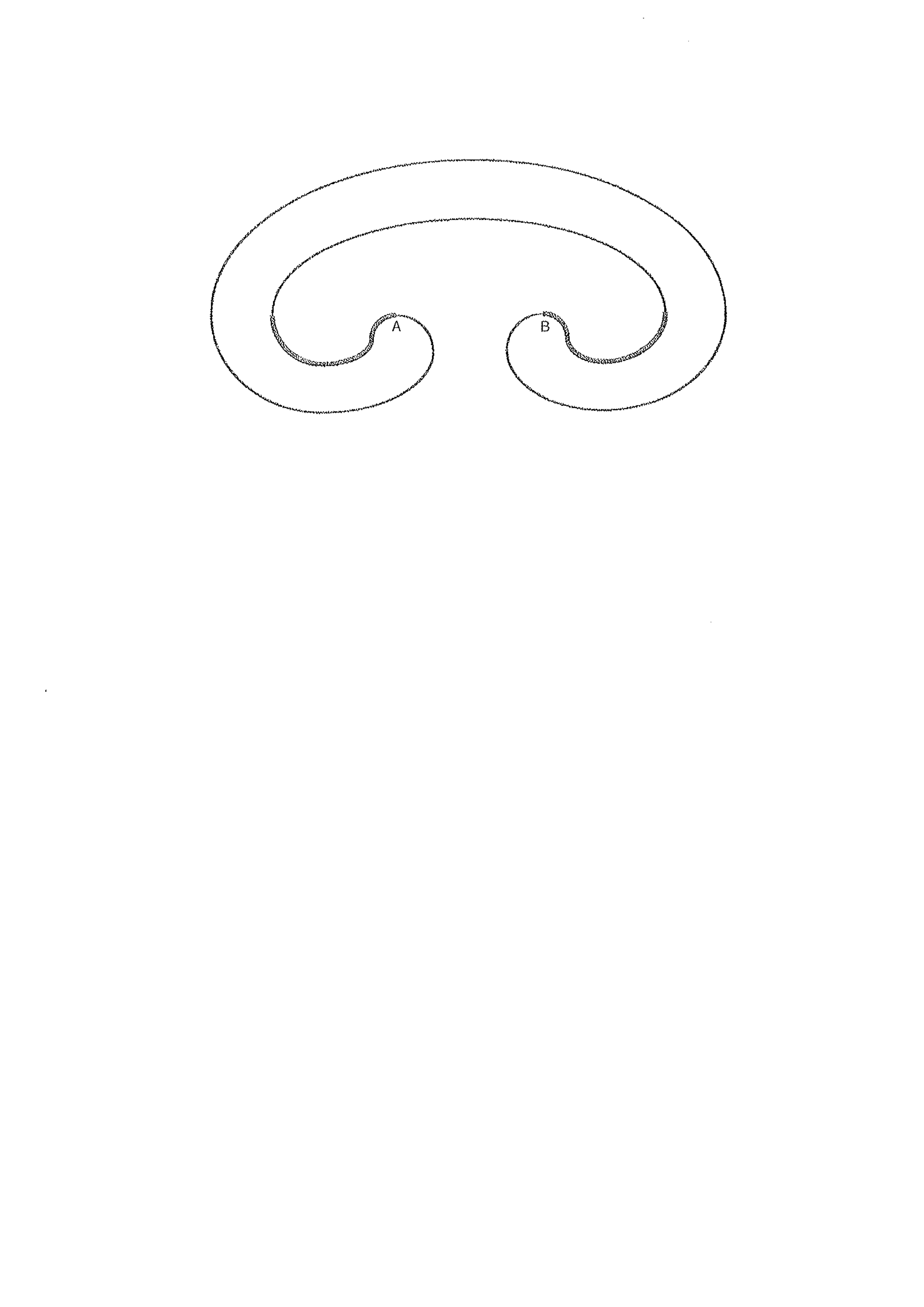}

\noindent
{\bf Figure 1}:
{\footnotesize Livshits' Example (adapted from Ch. 5 of \cite{kn:M}):
the internal upper part of the figure is half an ellipse with foci $A$ and $B$. A ray entering the interior of the ellipse
between the foci must exit between the foci after reflection. So,
no scattering ray has a common point  with the bold parts of the boundary.}

\end{document}